# A peridynamic approach to flexoelectricity

Pranesh Roy, Debasish Roy[,*]

*Computational Mechanics Lab, Department of Civil Engineering, Indian Institute of Science, Bangalore 560012, India*

([*]*Corresponding author; email:* *royd@civil.iisc.ernet.in*)

## Abstract

A flexoelectric peridynamic (PD) theory is proposed. Using the PD framework, the formulation introduces, perhaps for the first time, a nanoscale flexoelectric coupling that entails non-uniform strain in centrosymmetric dielectrics. This potentially enables PD modeling of a large class of phenomena in solid dielectrics involving cracks, discontinuities etc. wherein large strain gradients are present and the classical electromechanical theory based on partial differential equations do not directly apply. PD electromechanical equations, derived from Hamilton's principle, are shown to satisfy global balance requirements. Linear PD constitutive equations reflect the electromechanical coupling effect, with the mechanical force state affected by the polarization state and the electrical force state in turn by the displacement state. An analytical solution of the PD electromechanical equations in the integral form is presented for the static case when a point mechanical force and a point electric force act in a three dimensional infinite solid dielectric. A parametric study on how the different length scales influence the response is also undertaken.

## I. INTRODUCTION

Electromechanical coupling phenomena have received considerable importance for a long time. Many dielectric materials exhibit electrical polarization under mechanical deformation and most of them also show the converse effect, i.e. mechanical deformation induced by an electrical field. These materials have familiar applications in smart structures as sensors and actuators, in transducers for energy harvesting, in microelectromechanical systems etc. Among them, piezoelectric materials have been extensively investigated and technologically exploited. Piezoelectricity is a coupling between strain and electrical polarization that, owing to crystallographic considerations, occurs in non-centrosymmetric systems only. On the other hand, flexoelectricity or reverse flexoelectricity, a coupling between strain gradient and polarization or between polarization gradient and strain respectively, exists in all dielectric materials including centrosymmetric systems, where high strain gradient locally breaks the inversion symmetry and induces polarization. Although flexoelectricity is a more general phenomenon than piezoelectricity, it has been long overlooked in solid dielectrics, perhaps because the effect is typically manifest only in nanoscales where high strain gradients are frequent. Thus with surging interest in engineering nano-structures, flexoelectricity has of late received widespread attention. Majdoub *et al.*[1,2] have shown that flexoelectricity can cause remarkable enhancement in the harvested power from lead zirconate titanate (PZT) and $\mathrm{BaTiO_3}$ nanobeams. Lee *et al.*[3] have measured nanoscale strain gradients in ferroelectric $\mathrm{HoMnO_3}$ epitaxial thin films, observed a marked flexoelectric effect and concluded that flexoelectricity may be used to tune the physical properties of these materials. Fousek *et al.*[4] have suggested a proper shaping of composite constituents so that very high strain gradients develop under an externally applied force thereby inducing polarization even if all the constituents are non-piezoelectric. Jiang *et al.*[5] have shown that flexoelectric nano-generators can significantly increase the harvested power. Biancoli *et al.*[6] have traced the high values of flexoelectric coefficients compared to theoretical predictions in polycrystalline, centrosymmetric perovskites e.g. $\mathrm{(Ba,Sr)TiO_3}$, to symmetry breaking in these materials as inhomogeneities develop under high-temperature processing. Cross *et al.*[7] have performed experiments to measure the flexoelectric coefficients for ferroelectric, incipient ferroelectric and relaxor ferroelectric perovskites. Zubko *et al.*[8] have calculated all components of flexoelectric tensor for single crystals of paraelectric $\mathrm{SrTiO_3}$. Flexoelectricity also exists in liquid crystals, mechano-sensitive bio-membranes and polymers. Breneman *et al.*[9] have investigated the origin of flexoelectricity for the biophysical mechanism in stereocilia of the inner ear. Apart from this, a theoretical framework for analyzing flexoelectricity phenomena in solids has also been presented (see Maranganti *et al.*[10], Sharma *et al.*[11], Majdoub*et et al.*[1-2], Resta[12], Chen[13], Stengel[14], Hong*et et al.*[15]).

Fragility of many flexoelectric materials and presence of voids and cracks reduce their serviceability when subjected to mechanical and/or electrical loading. The coupling between strain gradient and polarization or polarization gradient and strain poses even more complications in understanding the mechanics of fracture for these materials. As very high strain gradients tend to occur near the crack tip, flexoelectric coupling assumes significance in these regions. Recently Abdollahi *et al.*[16] have pointed out that a decrease in the structure size increases the fracture resistance and it turns out to be asymmetric with respect to the sign of polarization in flexoelectric materials. Principles underlying the fracture mechanics of piezoelectric and ferroelectric materials have been investigated to an extent (Zhang *et al.*[17], Kuna[18], Fang *et al.*[19]) and the adopted modeling approaches are mostly based on partial differential equations (PDEs) and the principle of local action. The PDE-based route however constitutes an unsuitable mathematical structure to deal with cracks, point defects or dislocations.

In departing from the PDE-centric setup, an interesting alternative would be to take recourse to the peridynamic (PD) theory of continuum mechanics (Silling[20]) which is generically suited to the mathematical modelling of structured continua involving discontinuities, e.g. crack nucleation and propagation. PD, by construction, also accounts for long range forces and thus qualifies as a theory for nonlocal continuum response. PD treats internal forces as a network of finite-distance interactions amongst a collection of material particles. It obtains the equation of motion in an integro-differential form instead of PDEs, thus permitting discontinuities as long as the spatial integrals remain Riemann integrable. A preliminary model, the bond based PD, envisions forces within each bond as dependent on the deformation of that bond only. The idea results in a restriction on the material behavior that can be represented; e.g. it forces Poisson's ratio to be 1/4 for isotropic materials. An amelioration of this inadequacy has been achieved through the state-based PD theory (Silling *et al.*[21]) according to which the bond force between two interacting particles is determined by the collective deformation of bonds within the horizon of a particle (the finite neighborhood over which it interacts with other particles), enabling it to model materials over the entire permissible range of Poisson's ratio.

In this work, a PD model for flexoelectricity is proposed. PD introduces length scales representing 'action at a distance' and thereby captures additional physical information which are essential in the nanoscale. An appreciation of how the response is modified owing to the combined effect of the PD length scale and the one associated with higher order gradient terms in the theory of flexoelectricity is an important aspect considered in this work. Starting with Hamilton's principle and writing Toupin's electric enthalpy density in the PD framework, PD electromechanical governing equations are derived. Global balance requirements are shown to be satisfied upon integrating these equations over the body domain. Considering small deformation and small polarization, linear PD constitutive equations are written. These equations are electromechanically coupled so that, just as the polarization state influences the mechanical force state, the displacement state affects the electrical force state. In the limit as the horizon size goes to zero, PD equations recover the classical electromechanical governing equations. On this basis, PD material parameters are derived in terms of the classical ones by writing the displacement, polarization and electric potential state vectors using Taylor's expansions, substituting them in the PD equilibrium equations whilst retaining terms up to appropriate orders and finally integrating and comparing coefficients with the classical governing equations. Using Fourier transforms, an analytical solution to PD electromechanical equations in the integral form is arrived at for the static case. An example is considered where a point mechanical force and a point electric force act in a three dimensional infinite solid dielectric. Effects of PD and gradient length scales on the response are also investigated.

The rest of the paper is arranged as follows. Section II provides a brief recap of the theories of flexoelectricity and PD. Section III gives a systematic derivation of the flexoelectric PD theory that includes Hamilton's principle, the governing equations, constitutive equations and PD material parameters in terms of the classical parameters. Section IV describes an analytical solution procedure and provides an integral representation of the solution. Section V contains illustrations based on an example problem and Section VI furnishes a few concluding remarks.

## II. THEORY OF FLEXOELECTRICITY AND PD

For completeness, the theory of flexoelectricity, governing equations and constitutive equations are briefly reviewed. A recap of the PD theory, including the equations of motion and constitutive equations, is also provided.

### A. Flexoelectricity theory

Following Maranganti *et al.*[10] and Majdoub *et al.*[1], a short account of the theory of flexoelectricity is given below. Consider the strictly static deformation of a flexoelectric body with volume $\mathcal{V}$ bounded by a surface $S$ in a vacuum $\mathcal{V}'$ so that $\mathcal{V} \cap \mathcal{V}' = \varphi$ and the total volume is $\mathcal{V}^* = \mathcal{V} \cup \mathcal{V}'$. Hamilton's principle in this case may be written as follows.

$$\delta\int_{t_1}^{t_2}\int_{\mathcal{V}^*}\left(-\mathrm{H}\right)dVdt+\int_{t_1}^{t_2}\int_{\mathcal{V}}\left(\mathbf{f}\cdot\delta\mathbf{u}+\mathbf{E}^0\cdot\delta\mathbf{P}\right)dVdt=0 \tag{1}$$

where $\mathrm{H}$ is Toupin's electric enthalpy density whose variation is given as

$$\begin{aligned}\delta\mathrm{H}&=\delta\mathrm{W}^L\left(\mathbf{e},\mathbf{P},\nabla\nabla\mathbf{u},\nabla\mathbf{P}\right)-\varepsilon_0\nabla\varphi\cdot\delta\nabla\varphi+\nabla\varphi\cdot\delta\mathbf{P}+\delta\nabla\varphi\cdot\mathbf{P}\\&=\delta\mathrm{W}^L+\mathbf{D}\cdot\delta\nabla\varphi+\nabla\varphi\cdot\delta\mathbf{P}\end{aligned} \tag{2}$$

$\mathrm{W}^L$ is the internal energy density, $\mathbf{D}=-\varepsilon_0\nabla\varphi+\mathbf{P}$ the electric displacement vector, $\mathbf{P}$ the polarization vector, $\varphi$ the potential of Maxwell's self-field $\mathbf{E}^{MS}=-\nabla\varphi$, $\mathbf{u}$ the displacement vector and $\mathbf{e}=\frac{1}{2}\left(\nabla\mathbf{u}+\left(\nabla\mathbf{u}\right)^T\right)$ the symmetric strain tensor. $\mathbf{f}$ and $\mathbf{E}^0$ are respectively the external body force and electric field.

Equilibrium equations, derivable from Eq. (1), are:

$$\nabla\cdot\boldsymbol{\sigma}+\mathbf{f}=0 \text{ in } \mathcal{V} \tag{3}$$

$$\bar{\mathbf{E}}+\nabla\cdot\tilde{\mathbf{E}}+\mathbf{E}^{MS}+\mathbf{E}^0=0 \text{ in } \mathcal{V} \tag{4}$$

$$-\varepsilon_0\nabla^2\varphi+\nabla\cdot\mathbf{P}=0 \text{ in } \mathcal{V} \quad \nabla^2\varphi=0 \text{ in } \mathcal{V}' \tag{5}$$

where $\boldsymbol{\sigma} = \mathbf{t} - \nabla \cdot \tilde{\mathbf{t}}$, $\mathbf{t}$ is the Cauchy stress, $\tilde{\mathbf{t}}$ a higher order stress tensor, $\bar{\mathbf{E}}$ the effective local electric force vector and $\tilde{\mathbf{E}}$ a higher order local electric force tensor.

For isotropic centrosymmetric flexoelectrics, the constitutive equations in component form may be written as

$$t_{ij} = \left(c_{12}\delta_{ij}\delta_{ps} + 2c_{44}\delta_{ip}\delta_{js}\right)e_{ps} + \left(d_{12}\delta_{ij}\delta_{ps} + d_{44}\left(\delta_{is}\delta_{jp} + \delta_{js}\delta_{ip}\right)\right)P_{p,s} \tag{6}$$

$$\begin{aligned} t_{ijm,m} = &\left(\left(c_{12} + 2c_{44}\right)l'^2\nabla^2\delta_{ij}\delta_{ps}\right)u_{p,s} + \left(f_{12}\delta_{pi}\delta_{js} + f_{44}\left(\delta_{ps}\delta_{ji} + \delta_{is}\delta_{jp}\right)\right)P_{p,s} \\ &+ c_{44}l^2\nabla^2\left(\delta_{is}\delta_{jp} - \delta_{js}\delta_{ip}\right)u_{p,s} \end{aligned} \tag{7}$$

$$E_{ij} = \left(d_{12}\delta_{ij}\delta_{ps} + d_{44}\left(\delta_{is}\delta_{jp} + \delta_{js}\delta_{ip}\right)\right)u_{p,s} + \left(b_{12}\delta_{ij}\delta_{ps} + \left(b_{44} + b_{77}\right)\delta_{is}\delta_{jp} + \left(b_{44} - b_{77}\right)\delta_{js}\delta_{ip}\right)P_{p,s} \tag{8}$$

$$E_i = -aP_i - \left(f_{12}\delta_{ij}\delta_{ps} + f_{44}\left(\delta_{is}\delta_{jp} + \delta_{js}\delta_{ip}\right)\right)u_{j,ps} \tag{9}$$

where $l$ and $l'$ are the length scale parameters associated with the higher order gradient terms. On substituting the constitutive equations in the equilibrium equation, the following equations emerge.

$$\begin{aligned} &c_{44}\nabla^2\mathbf{u} + \left(c_{12} + c_{44}\right)\nabla\nabla\cdot\mathbf{u} - \left(c_{12} + 2c_{44}\right)l'^2\nabla^2\nabla\nabla\cdot\mathbf{u} - c_{44}l^2\left(\nabla^2\nabla^2\mathbf{u} - \nabla^2\nabla\nabla\cdot\mathbf{u}\right) \\ &+ \left(d_{44} - f_{12}\right)\nabla^2\mathbf{P} + \left(d_{12} + d_{44} - 2f_{44}\right)\nabla\nabla\cdot\mathbf{P} + \mathbf{f} = 0 \end{aligned} \tag{10}$$

$$\begin{aligned} &\left(d_{44} - f_{12}\right)\nabla^2\mathbf{u} + \left(d_{12} + d_{44} - 2f_{44}\right)\nabla\nabla\cdot\mathbf{u} \\ &+ \left(b_{44} + b_{77}\right)\nabla^2\mathbf{P} + \left(b_{12} + b_{44} - b_{77}\right)\nabla\nabla\cdot\mathbf{P} - a\mathbf{P} - \nabla\varphi + \mathbf{E}^0 = 0 \end{aligned} \tag{11}$$

$$-\varepsilon_0\nabla^2\varphi + \nabla\cdot\mathbf{P} = 0 \tag{12}$$

### B. PD theory

A brief overview of state-based and bond-based PD theories is presented following the approach by Silling *et al.*[20,21]. Considering static deformation, equilibrium equations for the state-based PD are given as

$$\int_{\mathcal{H}(\mathbf{x})}\left\{\underline{\mathbf{T}}[\mathbf{x}]\langle\xi\rangle - \underline{\mathbf{T}}[\mathbf{x}+\xi]\langle-\xi\rangle\right\}dV_{\mathbf{x}'} + \mathbf{b}(\mathbf{x}) = \mathbf{0} \tag{13}$$

where $\mathbf{x}$ is the position vector of a particle in the reference configuration $\mathcal{B}_0 \subset \mathbb{R}^3$, $\mathbf{x}'$ the position vector of a neighbouring particle in $\mathcal{B}_0$, $\xi = \mathbf{x}' - \mathbf{x}$ the bond vector, $\underline{\mathbf{T}}$ the force vector state field and $\mathbf{b}$ the body force density field. $\mathcal{H}(\mathbf{x})$ denotes the horizon and is defined by $\mathcal{H}(\mathbf{x}) = \left\{ \boldsymbol{\xi} \in \mathbb{R}^3 \mid (\boldsymbol{\xi}+\mathbf{x}) \in \mathcal{B}_0, |\boldsymbol{\xi}| < \delta \right\}$, where $\delta$ is the radius of the horizon.

On the other hand, for the bond-based PD, the static equilibrium equation takes the following form.

$$\int_{\mathcal{H}(\mathbf{x})} \mathbf{f}\left(\mathbf{u}(\mathbf{x}') - \mathbf{u}(\mathbf{x}), \mathbf{x}' - \mathbf{x}\right) dV_{\mathbf{x}'} + \mathbf{b}(\mathbf{x}) = 0 \tag{14}$$

where $\mathbf{f}$ is the force that the particle $\mathbf{x}'$ exerts on $\mathbf{x}$ and $\mathbf{u}$ is the displacement field. Conservation of angular momentum implies the following restriction on $\underline{\mathbf{T}}$ and $\mathbf{f}$.

$$\int_{\mathcal{H}(\mathbf{x})} \underline{\mathbf{T}}[\mathbf{x}]\langle \xi \rangle \times \underline{\mathbf{Y}}[\mathbf{x}]\langle \xi \rangle \, dV_{\mathrm{X}'} = 0, \quad \forall \mathbf{x} \in \mathcal{B}_0 \tag{15}$$

$$(\boldsymbol{\xi} + \boldsymbol{\eta}) \times \mathbf{f}(\boldsymbol{\eta}, \boldsymbol{\xi}) = 0 \qquad \forall \boldsymbol{\eta}, \boldsymbol{\xi} \tag{16}$$

where $\underline{\mathbf{Y}}[\mathbf{x}]\langle \xi \rangle = \mathbf{y}' - \mathbf{y} = \chi(\mathbf{x}') - \chi(\mathbf{x})$ is the deformed bond vector under the deformation map $\chi : \mathcal{B}_o \to \mathcal{B}_t$ from $\mathcal{B}_o$ to the current configuration $\mathcal{B}_t$ and $\boldsymbol{\eta} = \mathbf{u}(\mathbf{x}') - \mathbf{u}(\mathbf{x})$ the relative displacement.

Considering small deformations, $|\boldsymbol{\eta}| \ll 1$, constitutive equations for the bond-based PD may be written in the form (Silling *et al.*[20,22], Weckner *et al.*[23])

$$\mathbf{f}(\boldsymbol{\eta}, \boldsymbol{\xi}) = \Lambda(|\boldsymbol{\xi}|)(\boldsymbol{\xi} \otimes \boldsymbol{\xi}) \cdot \left(\mathbf{u}(\mathbf{x}') - \mathbf{u}(\mathbf{x})\right) = \mathbf{C}(\boldsymbol{\xi}) \cdot (\mathbf{u}' - \mathbf{u}) \tag{17}$$

$\mathbf{C}(\boldsymbol{\xi})$ is the symmetric micromodulus tensor and displacement vectors evaluated at $\mathbf{x}$ and $\mathbf{x}'$ are denoted as $\mathbf{u}$ and $\mathbf{u}'$ respectively. Substituting the constitutive equation in the equilibrium equation, the following can be written.

$$\int_{\mathcal{H}(\mathbf{x})} \mathbf{C}(\boldsymbol{\xi}) \cdot (\mathbf{u}' - \mathbf{u}) \, dV' + \mathbf{b}(\mathbf{x}) = 0 \tag{18}$$

## III. A FLEXOELECTRIC PD THEORY

The stage is now set for a flexoelectric PD theory - the focus of this work. We assume the existence of a mechanical force state, electrical force state and electric displacement state in response to externally applied mechanical body force and electric field. The following vector states are introduced to describe the state of the flexoelectric body.

A relative displacement vector state defined as:

$$\underline{\mathbf{U}}[\mathbf{x}]\langle\xi\rangle = \mathbf{u}' - \mathbf{u} \tag{19}$$

A relative polarization vector state of the form:

$$\underline{\mathbf{P}}[\mathbf{x}]\langle\xi\rangle = \mathbf{P}' - \mathbf{P} \tag{20}$$

A relative electric potential scalar state: $\underline{\Phi}[\mathbf{x}]\langle\xi\rangle = \varphi' - \varphi$ (21)

### A. Governing equations by Hamilton's principle

For a deformable solid flexoelectric body, under static deformation, Hamilton's principle may be stated as

$$\delta\int_{t_1}^{t_2}\int_{\mathcal{V}^*}(-\mathrm{H})\,dVdt + \int_{t_1}^{t_2}\int_{\mathcal{V}}\left(\mathbf{f}\cdot\delta\mathbf{u} + \mathbf{E}^0\cdot\delta\mathbf{P}\right)dVdt = 0 \tag{22}$$

Variation of Toupin's electric enthalpy density in the PD framework is proposed in the following way.

$$\delta\mathrm{H} = \delta w^L + \underline{\mathrm{d}}\bullet\delta\underline{\Phi} - \mathbf{E}^{MS}\cdot\delta\mathbf{P} \tag{23}$$

where $\underline{\mathrm{d}}$ is the electric displacement scalar state which is conjugate to the electric potential scalar state $\underline{\Phi}$ and $\bullet$ indicates the inner product $\left(\underline{\mathbf{A}}\bullet\underline{\mathbf{B}} = \int_{\mathcal{H}(\mathbf{x})}\underline{\mathbf{A}}\langle\xi\rangle\cdot\underline{\mathbf{B}}\langle\xi\rangle\,dV_{\mathbf{x}'}\right)$. Allowing the PD internal energy density $w^L$ to depend on $\underline{\mathbf{U}}$, $\underline{\mathbf{P}}$ and polarization vector $\mathbf{P}$ and assuming their independent variations, one arrives at the following identity.

$$\begin{aligned}\delta\mathrm{H} &= \mathcal{D}_{\underline{\mathbf{U}}}w^L\bullet\delta\underline{\mathbf{U}} + \mathcal{D}_{\underline{\mathbf{P}}}w^L\bullet\delta\underline{\mathbf{P}} + \mathcal{D}_{\mathbf{P}}w^L\bullet\delta\mathbf{P} + \underline{\mathrm{d}}\bullet\delta\underline{\Phi} - \mathbf{E}^{MS}\cdot\delta\mathbf{P}\\ &= \underline{\mathbf{T}}^p\bullet\delta\underline{\mathbf{U}} + \underline{\tilde{\mathbf{E}}}^p\bullet\delta\underline{\mathbf{P}} - \bar{\mathbf{E}}\cdot\delta\mathbf{P} + \underline{\mathrm{d}}\bullet\delta\underline{\Phi} - \mathbf{E}^{MS}\cdot\delta\mathbf{P}\\ &= \underline{\mathbf{T}}^p\bullet\delta\underline{\mathbf{U}} + \underline{\tilde{\mathbf{E}}}^p\bullet\delta\underline{\mathbf{P}} - \left(\bar{\mathbf{E}} + \mathbf{E}^{MS}\right)\cdot\delta\mathbf{P} + \underline{\mathrm{d}}\bullet\delta\underline{\Phi}\end{aligned} \tag{24}$$

where $\mathcal{D}_{\underline{\mathbf{U}}}$, $\mathcal{D}_{\underline{\mathbf{P}}}$ and $\mathcal{D}_{\mathbf{P}}$ designate Fréchet derivatives with respect to $\underline{\mathbf{U}}$, $\underline{\mathbf{P}}$ and $\mathbf{P}$ respectively. $\underline{\mathbf{T}}^p = \mathcal{D}_{\underline{\mathbf{U}}}w^L$ is the mechanical force vector state, $\underline{\tilde{\mathbf{E}}}^p = \mathcal{D}_{\underline{\mathbf{P}}}w^L$ the electrical force vector state, $\bar{\mathbf{E}} = -\mathcal{D}_{\mathbf{P}}w^L$ the effective local electric force.

Substituting Eq. (24) into Eq. (22), the following identity is obtainable.

$$\begin{aligned}&-\int_{t_1}^{t_2}\int_{\mathcal{V}^*}\int_{\mathcal{H}(\mathbf{x})}\left(\underline{\mathbf{T}}^p\cdot\delta(\mathbf{u}'-\mathbf{u}) + \underline{\tilde{\mathbf{E}}}^p\cdot\delta(\mathbf{P}'-\mathbf{P}) - \left(\bar{\mathbf{E}} + \mathbf{E}^{MS}\right)\cdot\delta\mathbf{P} + \underline{\mathrm{d}}\cdot\delta(\varphi'\text{-}\varphi)\right)dV'dVdt\\ &+\int_{t_1}^{t_2}\int_{\mathcal{V}}\left(\mathbf{f}\cdot\delta\mathbf{u} + \mathbf{E}^0\cdot\delta\mathbf{P}\right)dVdt = 0\end{aligned} \tag{25}$$

Applying Fubini's theorem to the first, second and fourth terms in Eq. (25), we have:

$$\begin{aligned}&\int_{t_1}^{t_2}\int_{\mathcal{V}^*}\left\{-\int_{\mathcal{H}(\mathbf{x})}\left(\underline{\mathbf{T}}^{p'}-\underline{\mathbf{T}}^{p}\right)dV'+\mathbf{f}\right\}\cdot\delta\mathbf{u}\,dVdt\\&+\int_{t_1}^{t_2}\int_{\mathcal{V}^*}\left\{-\int_{\mathcal{H}(\mathbf{x})}\left(\underline{\tilde{\mathbf{E}}}^{p'}-\underline{\tilde{\mathbf{E}}}^{p}\right)dV'+\mathbf{E}^{MS}+\overline{\mathbf{E}}+\mathbf{E}^{0}\right\}\cdot\delta\mathbf{P}\,dVdt-\int_{t_1}^{t_2}\int_{\mathcal{V}^*}\int_{\mathcal{H}(\mathbf{x})}\left(\underline{\mathrm{d}}'-\underline{\mathrm{d}}\right)dV'\cdot\delta\varphi\,dVdt=0\end{aligned}\tag{26}$$

$\delta\mathbf{u}$, $\delta\mathbf{P}$ and $\delta\varphi$ being independent and arbitrary, the following equations of equilibrium are readily obtained.

$$\int_{\mathcal{H}(\mathbf{x})}\left\{\underline{\mathbf{T}}^{p}-\underline{\mathbf{T}}^{p'}\right\}dV'+\mathbf{f}=0 \quad \text{in } \mathcal{V}\tag{27}$$

$$\int_{\mathcal{H}(\mathbf{x})}\left\{\underline{\tilde{\mathbf{E}}}^{p}-\underline{\tilde{\mathbf{E}}}^{p'}\right\}dV'+\mathbf{E}^{MS}+\overline{\mathbf{E}}+\mathbf{E}^{0}=0 \quad \text{in } \mathcal{V}\tag{28}$$

$$\int_{\mathcal{H}(\mathbf{x})}\left(\underline{\mathrm{d}}-\underline{\mathrm{d}}'\right)dV'=0 \text{ in } \mathcal{V} \text{ and } \mathcal{V}'\tag{29}$$

**Proposition 3.1.** Let $\mathcal{V}$ *be a bounded body subjected to a body force field* $\mathbf{f}$ *and an electrical field* $\mathbf{E}^0$. *Let the mechanical force vector state field be* $\underline{\mathbf{T}}^p$, *the electrical force vector state field be* $\underline{\tilde{\mathbf{E}}}^p$ *and the scalar electrical displacement state field be* $\underline{\mathrm{d}}$. *Then if Eq.* (27), (28) *and* (29) *hold in* $\mathcal{V}$, *global balance laws for the case of static deformation are satisfied.*

**Proof.** Integrating Eq. (27), (28) and (29) over the entire body domain, the following equations may be written.

$$\int_{\mathcal{V}}\left\{\int_{\mathcal{H}(\mathbf{x})}\left\{\underline{\mathbf{T}}^{p}-\underline{\mathbf{T}}^{p'}\right\}dV'+\mathbf{f}\right\}dV=0\tag{30}$$

$$\int_{\mathcal{V}}\left\{\int_{\mathcal{H}(\mathbf{x})}\left\{\underline{\tilde{\mathbf{E}}}^{p}-\underline{\tilde{\mathbf{E}}}^{p'}\right\}dV'+\mathbf{E}^{MS}+\overline{\mathbf{E}}+\mathbf{E}^{0}\right\}dV=0\tag{31}$$

$$\int_{\mathcal{V}}\left(\int_{\mathcal{H}(\mathbf{x})}\left(\underline{\mathrm{d}}-\underline{\mathrm{d}}'\right)dV'\right)dV=0\tag{32}$$

All the inner integrals may be trivially extended from the horizon $\mathcal{H}$ to the body $\mathcal{V}$, as particles do not interact beyond the horizon. Using a change of variables $\mathbf{x}\to\mathbf{x}'$ and applying Fubini's theorem, following simplifications may be made.

$$\int_{\mathcal{V}}\int_{\mathcal{H}(\mathbf{x})}\left\{\underline{\mathbf{T}}^{p}-\underline{\mathbf{T}}^{p'}\right\}dV'dV=\int_{\mathcal{V}}\int_{\mathcal{V}}\left\{\underline{\mathbf{T}}^{p}-\underline{\mathbf{T}}^{p'}\right\}dV'dV=\int_{\mathcal{V}}\int_{\mathcal{V}}\left\{\underline{\mathbf{T}}^{p'}-\underline{\mathbf{T}}^{p}\right\}dVdV'=0 \tag{33}$$

$$\int_{\mathcal{V}}\int_{\mathcal{H}(\mathbf{x})}\left\{\tilde{\underline{\mathbf{E}}}^{p}-\tilde{\underline{\mathbf{E}}}^{p'}\right\}dV'dV=\int_{\mathcal{V}}\int_{\mathcal{V}}\left\{\tilde{\underline{\mathbf{E}}}^{p}-\tilde{\underline{\mathbf{E}}}^{p'}\right\}dV'dV=\int_{\mathcal{V}}\int_{\mathcal{V}}\left\{\tilde{\underline{\mathbf{E}}}^{p'}-\tilde{\underline{\mathbf{E}}}^{p}\right\}dVdV'=0 \tag{34}$$

$$\int_{\mathcal{V}}\int_{\mathcal{H}(\mathbf{x})}\left(\underline{\mathrm{d}}-\underline{\mathrm{d}}'\right)dV'dV=\int_{\mathcal{V}}\int_{\mathcal{V}}\left(\underline{\mathrm{d}}-\underline{\mathrm{d}}'\right)dV'dV=\int_{\mathcal{V}}\int_{\mathcal{V}}\left(\underline{\mathrm{d}}'-\underline{\mathrm{d}}\right)dVdV'=0 \tag{35}$$

Using Eq. (33) to (35), Eq. (30) to (32) may be written as

$$\int_{\mathcal{V}}\mathbf{f}\,dV=0 \tag{36}$$

$$\int_{\mathcal{V}}\left\{\mathbf{E}^{MS}+\bar{\mathbf{E}}+\mathbf{E}^{0}\right\}dV=0 \tag{37}$$

Eq. (32) is trivially satisfied as no free charge is considered. Eq. (36) and (37) are statements of global balance laws. This completes the proof. □

## B. Constitutive relations

Following the bond-based PD route and assuming small deformation and polarization, linear PD electromechanical constitutive equations are of the form (see Silling *et al.*[20,22], Weckner *et al.*[23]):

$$\begin{aligned}\underline{\mathbf{T}}^{p}&=\Lambda_{1}\left(|\xi|\right)\left(\xi\otimes\xi\right)\cdot\underline{\mathbf{U}}+\Lambda_{2}\left(|\xi|\right)\left(\xi\otimes\xi\right)\cdot\underline{\mathbf{P}}\\&=\mathrm{C}_{1}\left(\xi\right)\cdot\left(\mathbf{u}'-\mathbf{u}\right)+\mathrm{C}_{2}\left(\xi\right)\cdot\left(\mathbf{P}'-\mathbf{P}\right)\end{aligned} \tag{38}$$

$$\begin{aligned}\tilde{\underline{\mathbf{E}}}^{p}&=\Lambda_{3}\left(|\xi|\right)\left(\xi\otimes\xi\right)\cdot\underline{\mathbf{U}}+\Lambda_{4}\left(|\xi|\right)\left(\xi\otimes\xi\right)\cdot\underline{\mathbf{P}}\\&=\mathrm{C}_{3}\left(\xi\right)\cdot\left(\mathbf{u}'-\mathbf{u}\right)+\mathrm{C}_{4}\left(\xi\right)\cdot\left(\mathbf{P}'-\mathbf{P}\right)\end{aligned} \tag{39}$$

$$\bar{\mathbf{E}}=\int_{\mathcal{H}(\mathbf{x})}\Lambda_{5}\left(|\xi|\right)\left(\xi\otimes\xi\right)\cdot\underline{\mathbf{U}}dV'-a\mathbf{P}=\int_{\mathcal{H}(\mathbf{x})}\mathrm{C}_{5}\left(\xi\right)\cdot\left(\mathbf{u}'-\mathbf{u}\right)dV'-a\mathbf{P} \tag{40}$$

$$\mathbf{E}^{MS}=\int_{\mathcal{H}(\mathbf{x})}\Lambda_{6}\left(|\xi|\right)\underline{\Phi}\,\xi\,dV'=\int_{\mathcal{H}(\mathbf{x})}\mathrm{C}_{6}\left(\xi\right)\left(\varphi'-\varphi\right)dV' \tag{41}$$

$$\underline{\mathrm{d}}=\Lambda_{7}\left(|\xi|\right)\underline{\Phi}+\Lambda_{8}\left(|\xi|\right)\mathbf{P}\cdot\xi=\mathrm{C}_{7}\left(\xi\right)\left(\varphi'-\varphi\right)+\mathrm{C}_{8}\left(\xi\right)\cdot\mathbf{P} \tag{42}$$

where $\mathrm{C}_{n}\left(\xi\right)=\Lambda_{n}\left(|\xi|\right)\left(\xi\otimes\xi\right)$ $n=1,2,...,5$; $\mathrm{C}_{6}\left(\xi\right)=\Lambda_{6}\left(|\xi|\right)\xi$; $\mathrm{C}_{7}\left(\xi\right)=\Lambda_{7}\left(|\xi|\right)$; $\mathrm{C}_{8}\left(\xi\right)=\Lambda_{8}\left(|\xi|\right)\xi$ are micromodulus functions. An exponential form is assumed for all $\Lambda$-s, viz.

$$\Lambda_1\left(|\xi|\right)=\left(c_{11}^{p}e^{-\left(\frac{|\xi|}{\delta}\right)^2}+c_{12}^{p}e^{-\left(\frac{|\xi|}{\delta_1}\right)^2}\right);\quad \delta_1\neq\delta \tag{43}$$

$$\Lambda_n\left(|\xi|\right)=c_n^{p}e^{-\left(\frac{|\xi|}{\delta}\right)^2},\quad n=2,3,\ldots,8 \tag{44}$$

Here $\delta$ and $\delta_1$ are the PD length scales and they determine the degree of nonlocality.

## 1. PD material parameters

PD electromechanical governing equations recover the classical equations in the limit as the horizon size decreases to zero. Substituting PD constitutive equations, i.e. Eq. (38) to (42), in the PD equilibrium equations, i.e. Eq. (27) to (29), the following equations may be written.

$$2\int_{\mathcal{H}(\mathbf{x})}\mathrm{C}_1(\xi)\cdot(\mathbf{u}'-\mathbf{u})\,dV'+2\int_{\mathcal{H}(\mathbf{x})}\mathrm{C}_2(\xi)\cdot(\mathbf{P}'-\mathbf{P})\,dV'+\mathbf{f}=0 \tag{45}$$

$$2\int_{\mathcal{H}(\mathbf{x})}\tilde{\mathrm{C}}(\xi)\cdot(\mathbf{u}'-\mathbf{u})\,dV'+2\int_{\mathcal{H}(\mathbf{x})}\mathrm{C}_4(\xi)\cdot(\mathbf{P}'-\mathbf{P})\,dV'+\int_{\mathcal{H}(\mathbf{x})}\mathrm{C}_6(\xi)(\varphi'-\varphi)\,dV'-a\mathbf{P}+\mathbf{E}^0=0 \tag{46}$$

$$2\int_{\mathcal{H}(\mathbf{x})}\mathrm{C}_7(\xi)(\varphi'-\varphi)\,dV'+\int_{\mathcal{H}(\mathbf{x})}\mathrm{C}_8(\xi)\cdot(\mathbf{P}+\mathbf{P}')\,dV'=0 \tag{47}$$

where $2\tilde{\mathrm{C}}=2\mathrm{C}_3+\mathrm{C}_5$.

Following Taylor's expansions of $\underline{\mathbf{U}}\langle\xi\rangle$, $\underline{\mathbf{P}}\langle\xi\rangle$ and $\underline{\Phi}\langle\xi\rangle$ and substituting Eq. (43) and (44) into Eq. (45) to (47) whilst retaining terms up to appropriate orders (i.e. consistent with those appearing in Eq. (10) to (12)) and performing integration, the following equations finally result.

$$\begin{aligned}&\frac{4\pi}{15}\left(I_6c_{11}^{p}+I_6'c_{12}^{p}\right)\nabla^2\mathbf{u}+\frac{8\pi}{15}\left(I_6c_{11}^{p}+I_6'c_{12}^{p}\right)\nabla\nabla\cdot\mathbf{u}+\frac{\pi}{105}\left(I_8c_{11}^{p}+I_8'c_{12}^{p}\right)\nabla^2\nabla^2\mathbf{u}\\&+\frac{4\pi}{105}\left(I_8c_{11}^{p}+I_8'c_{12}^{p}\right)\nabla^2\nabla\nabla\cdot\mathbf{u}+\frac{4\pi}{15}I_6c_2^{p}\,\nabla^2\mathbf{P}+\frac{8\pi}{15}I_6c_2^{p}\,\nabla\nabla\cdot\mathbf{P}+\mathbf{f}=\mathbf{0}\end{aligned} \tag{48}$$

$$\frac{4\pi}{15}I_6\tilde{c}^{p}\nabla^2\mathbf{u}+\frac{8\pi}{15}I_6\tilde{c}^{p}\nabla\nabla\cdot\mathbf{u}+\frac{4\pi}{15}I_6c_4^{p}\,\nabla^2\mathbf{P}+\frac{8\pi}{15}I_6c_4^{p}\,\nabla\nabla\cdot\mathbf{P}+\frac{4\pi}{3}I_4c_6^{p}\nabla\varphi\text{-}a\mathrm{P}+\mathrm{E}^0=\mathbf{0} \tag{49}$$

$$\frac{4\pi}{3}I_4c_7^{p}\nabla^2\varphi+\frac{4\pi}{3}I_4c_8^{p}\nabla\cdot\mathbf{P}=\mathbf{0} \tag{50}$$

Equating coefficients of the independent variables in the PD [Eq. (48) to (50)] and classical [Eq. (10) to (12)] theories, the PD material parameters are expressed in terms of the classical ones as follows.

$$c_{11}^p = \frac{15}{4}\frac{c_{44}\left(28I_6'l^2 + I_8'\right)}{\pi\left(I_6I_8' - I_6'I_8\right)},\ c_{12}^p = -\frac{15}{4}\frac{c_{44}\left(28I_6l^2 + I_8\right)}{\pi\left(I_6I_8' - I_6'I_8\right)},\ c_2^p = \frac{15d^*}{4\pi I_6},\ \tilde{c}^p = \frac{15d^*}{4\pi I_6},\ c_4^p = \frac{15b^*}{4\pi I_6},$$
$$c_6^p = -\frac{3}{4\pi I_4},\ c_7^p = -\frac{3\varepsilon_0}{4\pi I_4},\ c_8^p = \frac{3}{4\pi I_4} \tag{51}$$

where $I_n = \int_0^\infty e^{-\left(\frac{r}{\delta}\right)^2} r^n dr = \begin{cases} \dfrac{(n-1)!!\delta^{n+1}\sqrt{\pi}}{2^{n/2+1}} & \text{for } n \text{ even} \\ \dfrac{\left[\frac{1}{2}(n-1)\right]!\delta^{n+1}}{2} & \text{for } n \text{ odd} \end{cases}$.

and

$$n!! = 2^k k! \quad \text{for even positive integer } n = 2k, k \geq 0$$
$$= \frac{(2k)!}{2^k k!} \quad \text{for odd positive integer } n = 2k-1, k \geq 1$$

As a bond-based PD approach is presently followed, certain restrictions on the classical material parameters arise, viz. $c_{44} = c_{12}$, $\left(\frac{l}{l'}\right)^2 = \frac{3}{5}$, $d^* = d_{44} - f_{12}$, $b^* = b_{44} + b_{77}$ with $d^*$ and $b^*$ treated as only two constants associated with $d_{44}$, $f_{12}$ and $b_{44}$, $b_{77}$ respectively.

## IV. ANALYTICAL SOLUTION

In this section, the PD electromechanical governing equations are analytically solved using direct and inverse Fourier transforms. We refer to Weckner *et al.*[23] for bond-based PD solutions. Solutions to flexoelectric equations are presented here in an integral form.

### A. PD governing equations in Fourier space

Taking Fourier transform of Eq. (45) to (47) with respect to spatial coordinate $\mathbf{x}$ results in the following equations.

$$2\mathbf{M}_1(\mathbf{k})\cdot\bar{\mathbf{u}}(\mathbf{k})+2\mathbf{M}_2(\mathbf{k})\cdot\bar{\mathbf{P}}(\mathbf{k})=\bar{\mathbf{f}}(\mathbf{k}) \tag{52}$$

$$2\tilde{\mathbf{M}}(\mathbf{k})\cdot\bar{\mathbf{u}}(\mathbf{k})+2\mathbf{M}_4(\mathbf{k})\cdot\bar{\mathbf{P}}(\mathbf{k})+\mathbf{M}_6(\mathbf{k})\bar{\varphi}(\mathbf{k})+a\bar{\mathbf{P}}(\mathbf{k})=\mathbf{E}^0(\mathbf{k}) \tag{53}$$

$$2\mathrm{M}_7(\mathbf{k})\bar{\varphi}(\mathbf{k})-\mathbf{M}_8(\mathbf{k})\cdot\bar{\mathbf{P}}(\mathbf{k})=\mathbf{0} \tag{54}$$

where

$$\begin{aligned}&\mathbf{M}_n(\mathbf{k})=\bar{\mathbf{C}}_n(\mathbf{0})-\bar{\mathbf{C}}_n(\mathbf{k}) \quad n=1,2,4\\ &\tilde{\mathbf{M}}(\mathbf{k})=\bar{\tilde{\mathbf{C}}}(\mathbf{0})-\bar{\tilde{\mathbf{C}}}(\mathbf{k});\ \mathbf{M}_6(\mathbf{k})=\bar{\mathbf{C}}_6(\mathbf{0})+\bar{\mathbf{C}}_6(\mathbf{k});\ \mathrm{M}_7(\mathbf{k})=\bar{\mathrm{C}}_7(\mathbf{0})-\bar{\mathrm{C}}_7(\mathbf{k});\\ &\mathbf{M}_8(\mathbf{k})=\bar{\mathbf{C}}_8(\mathbf{0})-\bar{\mathbf{C}}_8(\mathbf{k})\end{aligned} \tag{55}$$

As $\mathrm{C}_3$ and $\mathrm{C}_5$ are combined into $\tilde{\mathrm{C}}$ as $2\tilde{\mathrm{C}}=2\mathrm{C}_3+\mathrm{C}_5$, the subscripts 3 and 5 do not appear in Eq. (55). Upon further simplifications, $\mathbf{M}_n(\mathbf{k})$, $\tilde{\mathbf{M}}(\mathbf{k})$, $\mathbf{M}_6(\mathbf{k})$ and $\mathbf{M}_8(\mathbf{k})$ may be written in the following form.

1. $$\mathbf{M}_n(\mathbf{k})=(\mathrm{M}_n)_{\parallel}(k)\mathbf{n}_\mathbf{k}\otimes\mathbf{n}_\mathbf{k}+(\mathrm{M}_n)_{\perp}(k)\mathbf{P}_{\mathbf{n}_\mathbf{k}} \quad n=1,2,4 \tag{56}$$

$$(\mathrm{M}_n)_{\parallel}(k)=4\pi\int_0^\infty\Lambda_n(r)r^4\left(\frac{1}{3}-\frac{\sin kr}{kr}-\frac{2\cos kr}{(kr)^2}+\frac{2\sin kr}{(kr)^3}\right)dr \tag{57}$$

$$(\mathrm{M}_n)_{\perp}(k)=4\pi\int_0^\infty\Lambda_n(r)r^4\left(\frac{1}{3}+\frac{\cos kr}{(kr)^2}-\frac{\sin kr}{(kr)^3}\right)dr \tag{58}$$

2. $$\tilde{\mathbf{M}}(\mathbf{k})=\tilde{\mathrm{M}}_{\parallel}(k)\mathbf{n}_\mathbf{k}\otimes\mathbf{n}_\mathbf{k}+\tilde{\mathrm{M}}_{\perp}(k)\mathbf{P}_{\mathbf{n}_\mathbf{k}} \tag{59}$$

$$\tilde{\mathrm{M}}_{\parallel}(k)=2\pi\int_0^\infty\left(2\Lambda_3(r)+\Lambda_5(r)\right)r^4\left(\frac{1}{3}-\frac{\sin kr}{kr}-\frac{2\cos kr}{(kr)^2}+\frac{2\sin kr}{(kr)^3}\right)dr \tag{60}$$

$$\tilde{\mathrm{M}}_{\perp}(k)=2\pi\int_0^\infty\left(2\Lambda_3(r)+\Lambda_5(r)\right)r^4\left(\frac{1}{3}+\frac{\cos kr}{(kr)^2}-\frac{\sin kr}{(kr)^3}\right)dr \tag{61}$$

3. $$\mathbf{M}_6(\mathbf{k})=\mathrm{M}_6(k)\mathbf{n}_\mathbf{k} \tag{62}$$

$$\mathrm{M}_6(k)=-4\pi i\int_0^\infty\Lambda_6(r)r^3\left(\frac{\sin kr-kr\cos kr}{(kr)^2}\right)dr, \quad i=\sqrt{-1} \tag{63}$$

$$\mathrm{M}_7(k) = 4\pi\int_0^\infty \Lambda_7(r) r^2 \left(1 - \frac{\sin kr}{kr}\right) dr \tag{64}$$

4. $\mathbf{M}_8(\mathbf{k}) = \mathrm{M}_8(k)\mathbf{n_k}$ (65)

$$\mathrm{M}_8(k) = 4\pi i\int_0^\infty \Lambda_8(r) r^3 \left(\frac{\sin kr - kr\cos kr}{(kr)^2}\right) dr, \qquad i = \sqrt{-1} \tag{66}$$

$\mathbf{k}$ denotes the wave number vector, $k = \|\mathbf{k}\|$, $\mathbf{n_k}$ is the unit vector along $\mathbf{k}$ and $\mathbf{P}_{\mathbf{n_k}} = \mathbf{I} - \mathbf{n_k} \otimes \mathbf{n_k}$ is a projection operator. The subscripts || and $\perp$ bear the obvious meaning through their associations with the projection operators $\mathbf{n_k} \otimes \mathbf{n_k}$ and $\mathbf{P}_{\mathbf{n_k}}$ respectively.

**B. Solution**

From Eq. (54), the following can be written.

$$\overline{\varphi}(\mathbf{k}) = \frac{\mathbf{M}_8(\mathbf{k}) \cdot \overline{\mathbf{P}}(\mathbf{k})}{2\mathrm{M}_7(\mathbf{k})} \tag{67}$$

Substituting Eq. (67) into Eq. (53) and solving Eq. (52) and (53), one has

$$\begin{bmatrix} 2\mathbf{M}_1(\mathbf{k}) & 2\mathbf{M}_2(\mathbf{k}) \\ 2\tilde{\mathbf{M}}(\mathbf{k}) & 2\mathbf{M}_4(\mathbf{k}) + a\mathbf{I} + \dfrac{1}{2\mathrm{M}_7(\mathbf{k})}\mathbf{M}_6(\mathbf{k}) \otimes \mathbf{M}_8(\mathbf{k}) \end{bmatrix} \cdot \begin{Bmatrix} \overline{\mathbf{u}}(\mathbf{k}) \\ \overline{\mathbf{P}}(\mathbf{k}) \end{Bmatrix} = \begin{Bmatrix} \tilde{\mathbf{f}}(\mathbf{k}) \\ \overline{\mathbf{E}}^0(\mathbf{k}) \end{Bmatrix} \tag{68}$$

$$\begin{Bmatrix} \overline{\mathbf{u}}(\mathbf{k}) \\ \overline{\mathbf{P}}(\mathbf{k}) \end{Bmatrix} = \begin{bmatrix} \mathbf{M_{uf}} & \mathbf{M_{uE^0}} \\ \mathbf{M_{Pf}} & \mathbf{M_{PE^0}} \end{bmatrix} \cdot \begin{Bmatrix} \overline{\mathbf{f}}(\mathbf{k}) \\ \overline{\mathbf{E}}^0(\mathbf{k}) \end{Bmatrix} \tag{69}$$

where

$$\mathbf{M_{uf}} = \frac{2(\mathrm{M}_4)_{\|} + a + \frac{1}{2\mathrm{M}_7}\mathrm{M}_6\mathrm{M}_8}{2(\mathrm{M}_1)_{\|}\left(2(\mathrm{M}_4)_{\|} + \frac{1}{2\mathrm{M}_7}\mathrm{M}_6\mathrm{M}_8 + a\right) - 4\left(\tilde{M}\right)_{\|}(\mathrm{M}_2)_{\|}}\mathbf{n_k}\otimes\mathbf{n_k}$$

$$+\frac{2(\mathrm{M}_4)_{\perp} + a}{2(\mathrm{M}_1)_{\perp}\left(2(\mathrm{M}_4)_{\perp} + a\right) - 4\left(\tilde{\mathrm{M}}\right)_{\perp}(\mathrm{M}_2)_{\perp}}\mathbf{P_{n_k}} \tag{70}$$

$$= \left(\mathrm{M}_{\mathbf{uf}}\right)_{\|}\mathbf{n_k}\otimes\mathbf{n_k} + \left(\mathrm{M}_{\mathbf{uf}}\right)_{\perp}\mathbf{P_{n_k}}$$

$$\mathbf{M_{uE^0}} = \frac{-2(\mathrm{M}_2)_{\|}}{2(\mathrm{M}_1)_{\|}\left(2(\mathrm{M}_4)_{\|} + \frac{1}{2\mathrm{M}_7}\mathrm{M}_6\mathrm{M}_8 + a\right) - 4\left(\tilde{M}\right)_{\|}(\mathrm{M}_2)_{\|}}\mathbf{n_k}\otimes\mathbf{n_k}$$

$$+\frac{-2(\mathrm{M}_2)_{\perp}}{2(\mathrm{M}_1)_{\perp}\left(2(\mathrm{M}_4)_{\perp} + a\right) - 4\left(\tilde{\mathrm{M}}\right)_{\perp}(\mathrm{M}_2)_{\perp}}\mathbf{P_{n_k}} \tag{71}$$

$$= \left(\mathrm{M}_{\mathbf{uE^0}}\right)_{\|}\mathbf{n_k}\otimes\mathbf{n_k} + \left(\mathrm{M}_{\mathbf{uE^0}}\right)_{\perp}\mathbf{P_{n_k}}$$

$$\mathbf{M_{Pf}} = \frac{-2\left(\tilde{\mathrm{M}}\right)_{\|}}{2(\mathrm{M}_1)_{\|}\left(2(\mathrm{M}_4)_{\|} + \frac{1}{2\mathrm{M}_7}\mathrm{M}_6\mathrm{M}_8 + a\right) - 4\left(\tilde{M}\right)_{\|}(\mathrm{M}_2)_{\|}}\mathbf{n_k}\otimes\mathbf{n_k}$$

$$+\frac{-2\left(\tilde{\mathrm{M}}\right)_{\perp}}{2(\mathrm{M}_1)_{\perp}\left(2(\mathrm{M}_4)_{\perp} + a\right) - 4\left(\tilde{\mathrm{M}}\right)_{\perp}(\mathrm{M}_2)_{\perp}}\mathbf{P_{n_k}} \tag{72}$$

$$= \left(\mathrm{M}_{\mathbf{Pf}}\right)_{\|}\mathbf{n_k}\otimes\mathbf{n_k} + \left(\mathrm{M}_{\mathbf{Pf}}\right)_{\perp}\mathbf{P_{n_k}}$$

$$\mathbf{M_{PE^0}} = \frac{2(\mathrm{M}_1)_{\|}}{2(\mathrm{M}_1)_{\|}\left(2(\mathrm{M}_4)_{\|} + \frac{1}{2\mathrm{M}_7}\mathrm{M}_6\mathrm{M}_8 + a\right) - 4\left(\tilde{M}\right)_{\|}(\mathrm{M}_2)_{\|}}\mathbf{n_k}\otimes\mathbf{n_k}$$

$$+\frac{2(\mathrm{M}_1)_{\perp}}{2(\mathrm{M}_1)_{\perp}\left(2(\mathrm{M}_4)_{\perp} + a\right) - 4\left(\tilde{\mathrm{M}}\right)_{\perp}(\mathrm{M}_2)_{\perp}}\mathbf{P_{n_k}} \tag{73}$$

$$= \left(\mathrm{M}_{\mathbf{PE^0}}\right)_{\|}\mathbf{n_k}\otimes\mathbf{n_k} + \left(\mathrm{M}_{\mathbf{PE^0}}\right)_{\perp}\mathbf{P_{n_k}}$$

$\bar{\mathbf{u}}(\mathbf{k})$, $\bar{\mathbf{P}}(\mathbf{k})$ and $\bar{\varphi}(\mathbf{k})$ may be concisely expressed in the following forms.

$$\bar{\mathbf{u}}(\mathbf{k}) = \mathbf{M_{uf}}\cdot\bar{\mathbf{f}}(\mathbf{k}) + \mathbf{M_{uE^0}}\cdot\bar{\mathbf{E}}^0(\mathbf{k}) \tag{74}$$

$$\bar{\mathbf{P}}(\mathbf{k}) = \mathbf{M_{Pf}} \cdot \bar{\mathbf{f}}(\mathbf{k}) + \mathbf{M_{PE^0}} \cdot \bar{\mathbf{E}}^0(\mathbf{k}) \tag{75}$$

$$\begin{aligned}\bar{\varphi}(\mathbf{k}) &= \frac{1}{2\mathrm{M}_7(\mathbf{k})}\mathbf{M}_8(\mathbf{k}) \cdot \bar{\mathbf{P}}(\mathbf{k}) \\ &= \frac{1}{2\mathrm{M}_7(\mathbf{k})}\mathbf{M}_8(\mathbf{k}) \cdot \left(\mathbf{M_{Pf}} \cdot \bar{\mathbf{f}}(\mathbf{k}) + \mathbf{M_{uE^0}} \cdot \bar{\mathbf{E}}^0(\mathbf{k})\right) \\ &= \mathbf{M_{\varphi f}} \cdot \bar{\mathbf{f}}(\mathbf{k}) + \mathbf{M_{\varphi E^0}} \cdot \bar{\mathbf{E}}^0(\mathbf{k})\end{aligned} \tag{76}$$

Upon inverse transforms, one has:

$$\mathbf{u}(\mathbf{x}) = \mathcal{F}^{-1}\left\{\mathbf{M_{uf}}(\mathbf{k}) \cdot \bar{\mathbf{f}}(\mathbf{k}) + \mathbf{M_{uE^0}}(\mathbf{k}) \cdot \bar{\mathbf{E}}^0(\mathbf{k})\right\} \tag{77}$$

$$\mathbf{P}(\mathbf{x}) = \mathcal{F}^{-1}\left\{\mathbf{M_{Pf}}(\mathbf{k}) \cdot \bar{\mathbf{f}}(\mathbf{k}) + \mathbf{M_{PE^0}}(\mathbf{k}) \cdot \bar{\mathbf{E}}^0(\mathbf{k})\right\} \tag{78}$$

$$\varphi(\mathbf{x}) = \mathcal{F}^{-1}\left(\mathbf{M_{\varphi f}}(\mathbf{k}) \cdot \bar{\mathbf{f}}(\mathbf{k}) + \mathbf{M_{\varphi E^0}}(\mathbf{k}) \cdot \bar{\mathbf{E}}^0(\mathbf{k})\right) \tag{79}$$

It is usually non-trivial to find analytical solutions to Eq. (77) to (79) for a general geometry and loading configuration. In such cases, numerical techniques may be employed to get approximate solutions. In the following section, we consider a specific problem that admits a closed-form solution.

## V. AN ILLUSTRATIVE EXAMPLE

### A. A 3D infinite body under mechanical and electrical point forces

For demonstration purposes, a 3D infinite body under mechanical and electrical point forces given respectively by $\mathbf{f}(\mathbf{x}) = \tilde{\mathbf{f}}\delta(\mathbf{x})$ and $\mathbf{E^0}(\mathbf{x}) = \tilde{\mathbf{E}}^{\mathbf{0}}\delta(\mathbf{x})$ is taken; here $\delta(\mathbf{x})$ is the Dirac delta function. Considering interaction among all points and assuming micromodulus functions according to Eq. (43) and (44), we may simplify Eq. (57), (58), (60), (61), (63), (64) and (66) as follows.

$$(\mathrm{M}_1)_{||}(k) = \frac{\pi^{3/2}\delta^5 c_{11}^p}{4}\left(2 + e^{-\frac{k^2\delta^2}{4}}\left(k^2\delta^2 - 2\right)\right) + \frac{\pi^{3/2}\delta_1^5 c_{12}^p}{4}\left(2 + e^{-\frac{k^2\delta^2}{4}}\left(k^2\delta^2 - 2\right)\right) \tag{80}$$

$$\left(\mathrm{M}_1\right)_\perp(k)=\frac{\pi^{3/2}\delta^5 c_{11}^p}{2}\left(1-e^{-\frac{k^2\delta^2}{4}}\right)+\frac{\pi^{3/2}\delta_1^5 c_{12}^p}{2}\left(1-e^{-\frac{k^2\delta^2}{4}}\right) \tag{81}$$

$$\left(\mathrm{M}_n\right)_\parallel(k)=\frac{\pi^{3/2}\delta^5 c_n^p}{4}\left(2+e^{-\frac{k^2\delta^2}{4}}\left(k^2\delta^2-2\right)\right) \qquad n=2,4 \tag{82}$$

$$\left(\mathrm{M}_n\right)_\perp(k)=\frac{\pi^{3/2}\delta^5 c_n^p}{2}\left(1-e^{-\frac{k^2\delta^2}{4}}\right) \qquad n=2,4 \tag{83}$$

$$\left(\tilde{\mathrm{M}}\right)_\parallel(k)=\frac{\pi^{3/2}\delta^5 \tilde{c}^p}{4}\left(2+e^{-\frac{k^2\delta^2}{4}}\left(k^2\delta^2-2\right)\right) \tag{84}$$

$$\left(\tilde{\mathrm{M}}\right)_\perp(k)=\frac{\pi^{3/2}\delta^5 \tilde{c}^p}{2}\left(1-e^{-\frac{k^2\delta^2}{4}}\right) \tag{85}$$

$$\mathrm{M}_6(k)=-i\frac{\pi^{3/2}c_6^p\delta^5 k}{2}e^{-\frac{k^2\delta^2}{4}}, \qquad i=\sqrt{-1} \tag{86}$$

$$\mathrm{M}_7(k)=4\pi^{3/2}c_7^p\delta^3\left(1-e^{-\frac{k^2\delta^2}{4}}\right) \tag{87}$$

$$\mathrm{M}_8(k)=i\frac{\pi^{3/2}c_8^p\delta^5 k}{2}e^{-\frac{k^2\delta^2}{4}}, \qquad i=\sqrt{-1} \tag{88}$$

In this case, solutions to Eq. (77) to (79) may be expressed in the form as shown below.

$$\mathbf{u}(\mathbf{x})=\mathcal{F}^{-1}\left\{\mathbf{M}_{\mathbf{uf}}(\mathbf{k})\right\}\cdot\tilde{\mathbf{f}}+\mathcal{F}^{-1}\left\{\mathbf{M}_{\mathbf{uE^0}}(\mathbf{k})\right\}\cdot\tilde{\mathbf{E}}^0 \tag{89}$$

$$\mathbf{P}(\mathbf{x})=\mathcal{F}^{-1}\left\{\mathbf{M}_{\mathbf{Pf}}(\mathbf{k})\right\}\cdot\tilde{\mathbf{f}}+\mathcal{F}^{-1}\left\{\mathbf{M}_{\mathbf{PE^0}}(\mathbf{k})\right\}\cdot\tilde{\mathbf{E}}^0 \tag{90}$$

$$\varphi(\mathbf{x})=\mathcal{F}^{-1}\left(\mathbf{M}_{\varphi\mathbf{f}}(\mathbf{k})\right)\cdot\tilde{\mathbf{f}}+\mathcal{F}^{-1}\left(\mathbf{M}_{\varphi\mathbf{E^0}}(\mathbf{k})\right)\cdot\tilde{\mathbf{E}}^0 \tag{91}$$

For $\alpha=\mathbf{uf},\mathbf{uE^0},\mathbf{Pf},\mathbf{PE^0}$, we may write (see Weckner *et al.*[23]).

$$\mathcal{F}^{-1}\left\{\mathbf{M}_\alpha(\mathbf{k})\right\}=\left(\mathrm{f}_{\mathrm{n_x}}\right)_\alpha \mathbf{n_x}\otimes\mathbf{n_x}+\left(\mathrm{f}_{\mathrm{P_{n_x}}}\right)_\alpha(x)\mathbf{P_{n_x}} \tag{92}$$

where

$$\left(\mathrm{f}_{\mathrm{n}_{\mathrm{x}}}\right)_\alpha = \delta(\mathbf{x})\left(\mathrm{M}_\alpha\right)_\infty + \frac{1}{2\pi^2}\int_0^\infty \left[\left(-\frac{\sin kx}{kx} - \frac{2\cos kx}{(kx)^2} + \frac{2\sin kx}{(kx)^3}\right)\left(k^2\left(\mathrm{M}_\alpha\right)_\perp(k) - k^2\left(\mathrm{M}_\alpha\right)_\parallel(k)\right) + \frac{\sin kx}{kx}\left(k^2\left(\mathrm{M}_\alpha\right)_\perp(k) - k^2\left(\mathrm{M}_\alpha\right)_\infty\right)\right] dk \tag{93}$$

$$\left(\mathrm{f}_{\mathrm{P}_{\mathrm{n}_{\mathrm{x}}}}\right)_\alpha = \delta(\mathbf{x})\left(\mathrm{M}_\alpha\right)_\infty + \frac{1}{2\pi^2}\int_0^\infty \left[\left(\frac{\cos kx}{(kx)^2} - \frac{\sin kx}{(kx)^3}\right)\left(k^2\left(\mathrm{M}_\alpha\right)_\perp(k) - k^2\left(\mathrm{M}_\alpha\right)_\parallel(k)\right) + \frac{\sin kx}{kx}\left(k^2\left(\mathrm{M}_\alpha\right)_\perp(k) - k^2\left(\mathrm{M}_\alpha\right)_\infty\right)\right] dk \tag{94}$$

$\left(\mathrm{M}_\alpha\right)_\infty$ is $\mathrm{M}_\alpha$ at the small wavelength limit.

## B. Results

Borrowing from Maranganti *et al.*[10], the following material parameters are considered in this work:

$c_{44} = 0.325\times10^{-2}\ \mathrm{dyne/nm^2}$, $d_{44} = 0.356\times10^{15}\ \mathrm{dyne\,nm/C}$, $f_{12} = 0.01125\times10^{7}\ \mathrm{dyne\,nm/C}$, $b_{44} = 0.5255\times10^{32}\ \mathrm{dyne\,nm^4/C^2}$, $b_{77} = 1.921\times10^{32}\ \mathrm{dyne\,nm^4/C^2}$, $a = 8.767\times10^{33}\ \mathrm{dyne\,nm^2/C^2}$, $\varepsilon_0 = 8.854\times10^{-35}\ \mathrm{C^2/dyne\,nm^2}$, $\delta_1 = \dfrac{\delta}{2}$.

Responses are plotted for a point load $\left(\tilde{\mathbf{f}}\right)$ of magnitude 1 $\mathrm{dyne/nm^3}$ or an electric force $\left(\tilde{\mathbf{E}}^{\mathbf{0}}\right)$ of magnitude 1 dyne/C.

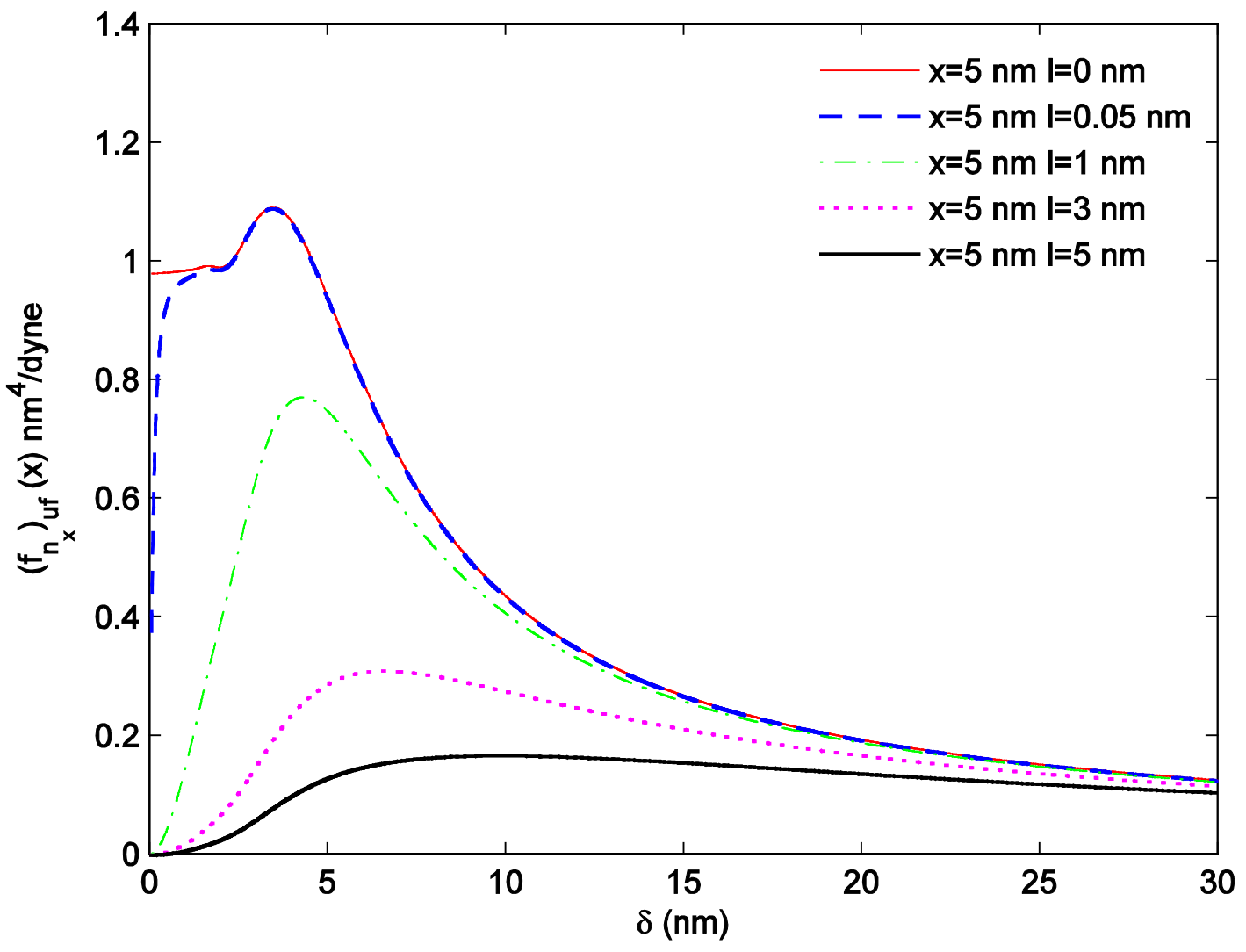

FIG. 1. Displacement $\left(f_{n_x}\right)_{\mathbf{uf}}$ for a unit static point load at a spatially fixed point with varying $\delta$ and $l$

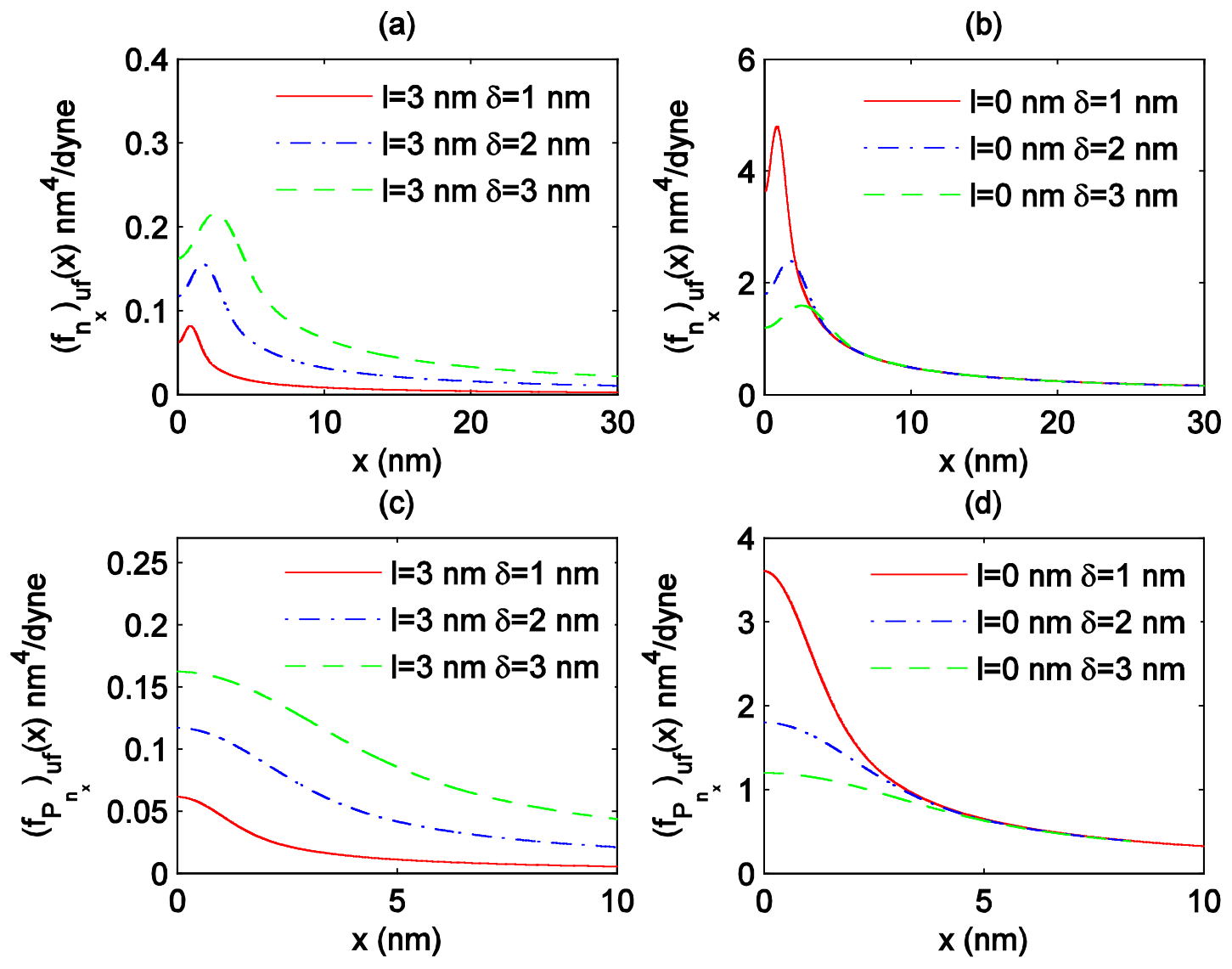

FIG. 2. (a) Displacement $\left(f_{n_x}\right)_{\mathbf{uf}}$ for a unit static point load for $l = 3\,\mathrm{nm}$. (b) Displacement $\left(f_{n_x}\right)_{\mathbf{uf}}$ for a unit static point load for $l = 0\,\mathrm{nm}$. (c) Displacement $\left(f_{P_{n_x}}\right)_{\mathbf{uf}}$ for a unit static point load for $l = 3\,\mathrm{nm}$. (d) Displacement $\left(f_{P_{n_x}}\right)_{\mathbf{uf}}$ for a unit static point load for $l = 0\,\mathrm{nm}$.

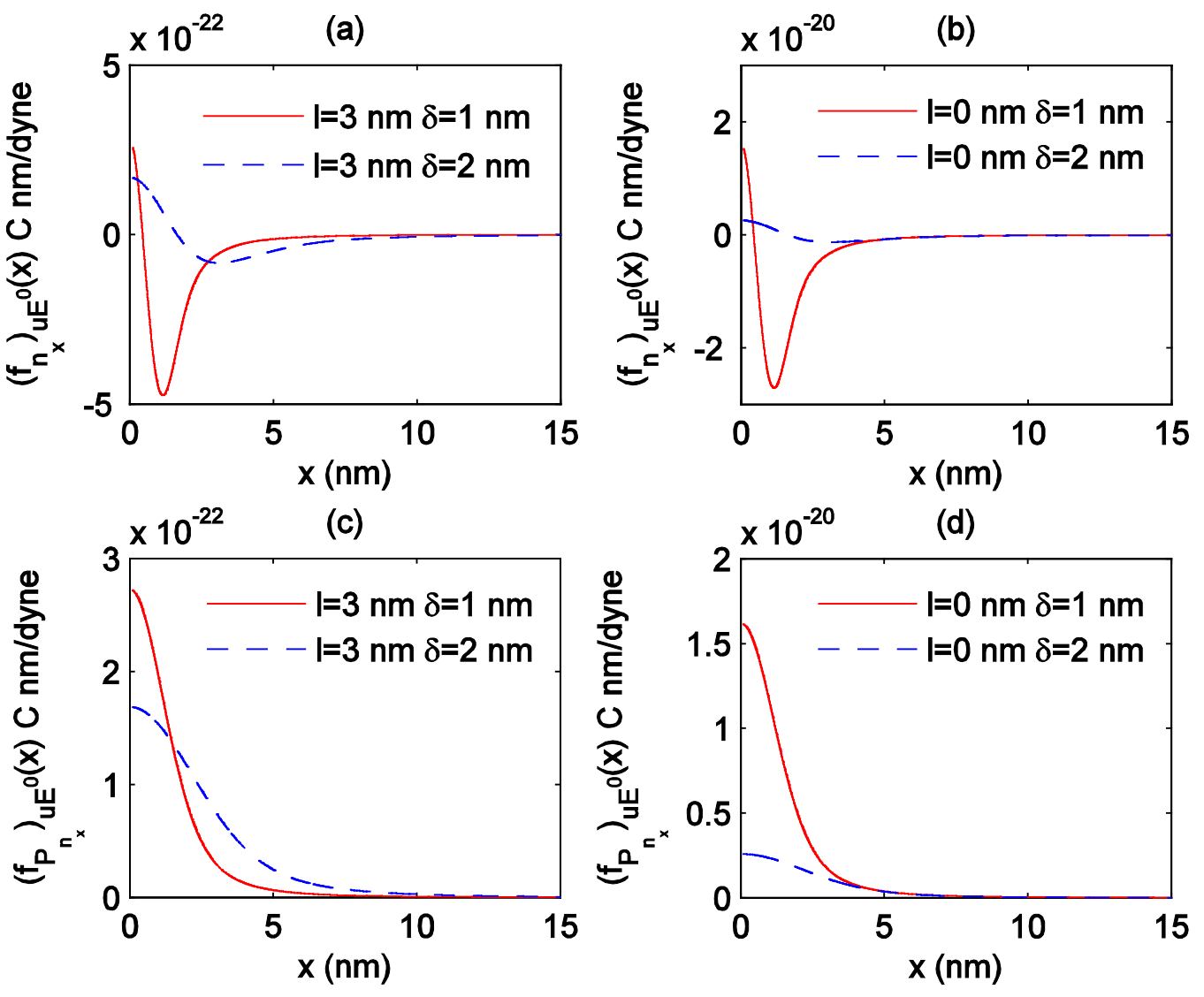

FIG. 3. (a) Displacement $\left(f_{n_x}\right)_{\mathbf{uE^0}}$ for a unit point electric force for $l=3\,\mathrm{nm}$. (b) Displacement $\left(f_{n_x}\right)_{\mathbf{uE^0}}$ for a unit point electric force for $l=0\,\mathrm{nm}$. (c) Displacement $\left(f_{P_{n_x}}\right)_{\mathbf{uE^0}}$ for a unit point electric force for $l=3\,\mathrm{nm}$. (d) Displacement $\left(f_{P_{n_x}}\right)_{\mathbf{uE^0}}$ for a unit point electric force for $l=0\,\mathrm{nm}$.

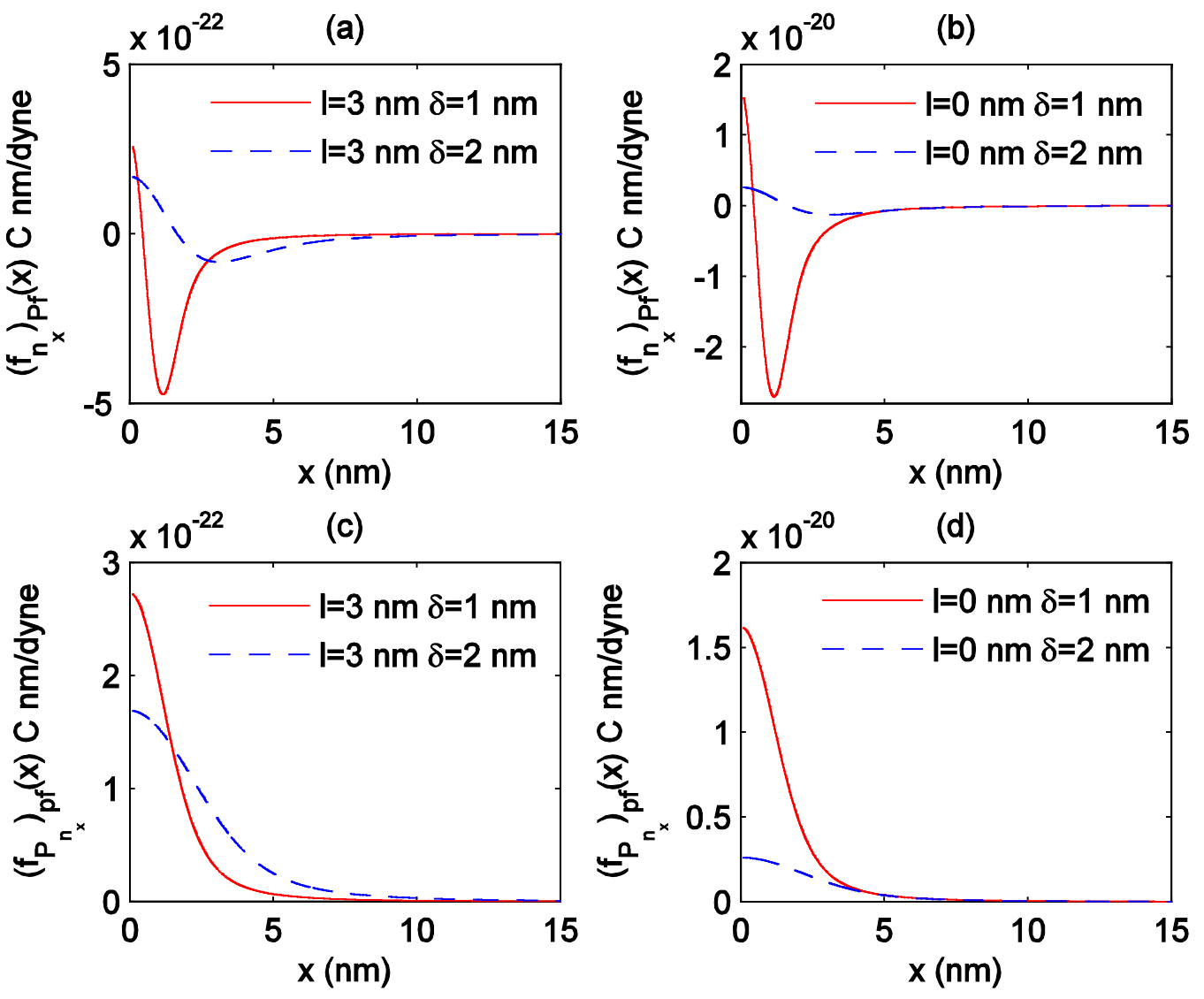

FIG. 4. (a) Polarization $\left(f_{n_x}\right)_{\mathbf{Pf}}$ for a unit static point load for $l=3\,\mathrm{nm}$. (b) Polarization $\left(f_{n_x}\right)_{\mathbf{Pf}}$ for a unit static point load for $l=0\,\mathrm{nm}$. (c) Polarization $\left(f_{P_{n_x}}\right)_{\mathbf{Pf}}$ for a unit static point load for $l=3\,\mathrm{nm}$. (d) Polarization $\left(f_{P_{n_x}}\right)_{\mathbf{Pf}}$ for a unit static point load for $l=0\,\mathrm{nm}$.

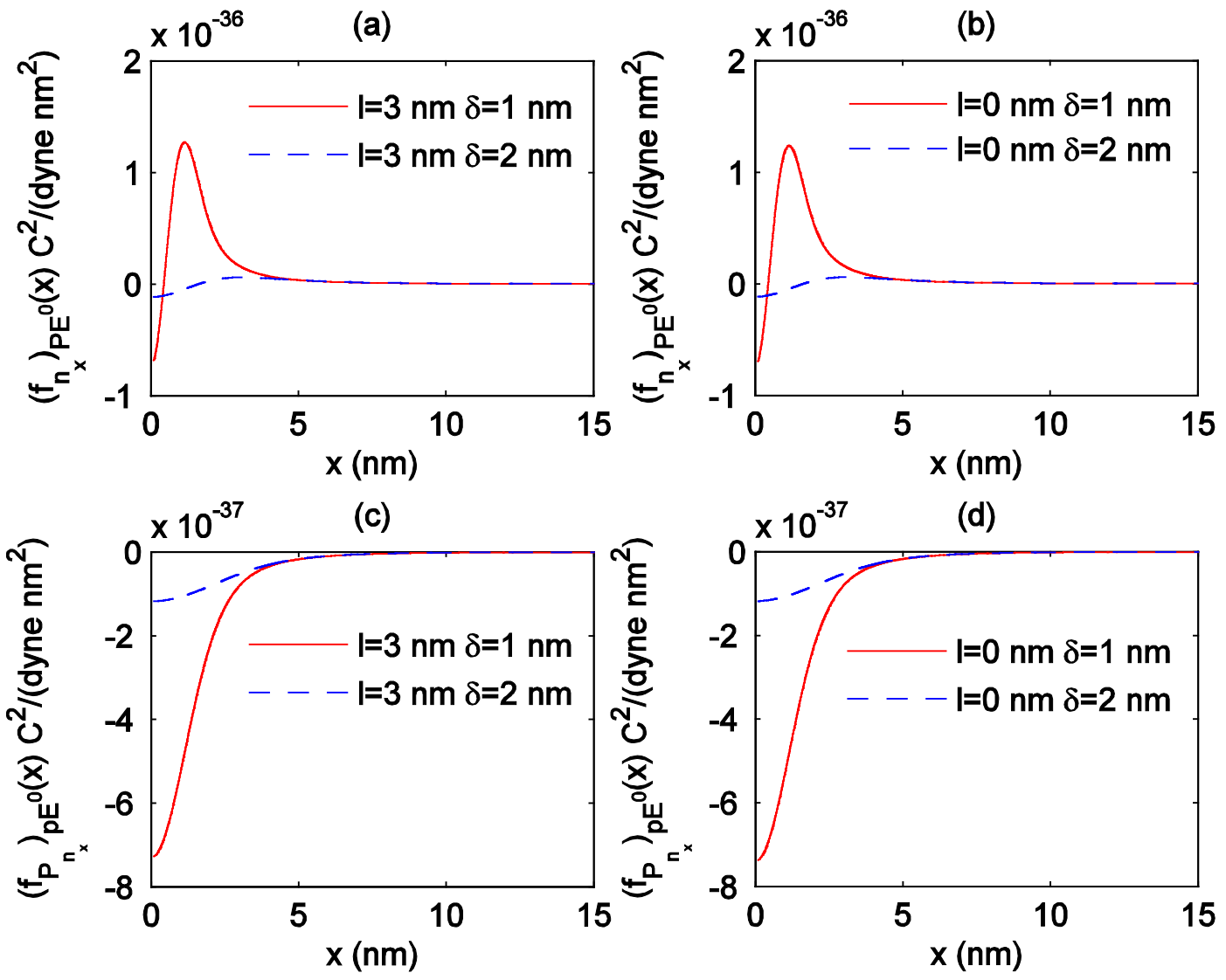

FIG. 5. (a) Polarization $\left(f_{n_x}\right)_{\mathbf{PE^0}}$ for a unit point electric force for $l = 3\,\mathrm{nm}$. (b) Polarization $\left(f_{n_x}\right)_{\mathbf{PE^0}}$ for a unit point electric force for $l = 0\,\mathrm{nm}$. (c) Polarization $\left(f_{P_{n_x}}\right)_{\mathbf{PE^0}}$ for a unit point electric force for $l = 3\,\mathrm{nm}$. (d) Polarization $\left(f_{P_{n_x}}\right)_{\mathbf{PE^0}}$ for a unit point electric force for $l = 0\,\mathrm{nm}$.

In Figure 1, when $\delta > x_{fixed}$ and $\forall l$, displacement decreases as $\delta$ increases, i.e. a stiffening of the response is observed with increasing $\delta$. When $\delta < x_{fixed}$, three regions are observed for $l = 0$. Here displacement initially remains almost constant, then increases (response softening) and after that decreases (stiffening) with increasing $\delta$. For non-zero $l's$ also, an initial increase is seen and, after reaching a peak, displacement decreases. An overall decrease in the displacement (stiffening) is observed with increasing $l$ for any $\delta$. In Figure 2, when $l = 0$ and as $\delta$ increases, a decrease in displacement (stiffening) is observed as we go away from the point of application of the load. But when $l = 3$, the opposite behaviour, i.e. softening with increasing $\delta$, is seen. It is concluded that the stiffening effect due to $l$ is more pronounced when $\delta$ is less. It may also be observed that the peak shifts to the right as $\delta$ increases. In Figures 3(a) and (b), displacement changes sign with increasing $x$, then becomes asymptotic to the $x$ axis. Curves become flatter with increasing $\delta$. In Figures 3(c) and (d), for $l = 0$, a stiffening behaviour is seen with increasing $\delta$ and for $l = 3$ two regions are observed where initial softening followed by a stiffening is noticed with increasing $x$. Figure 4 is similar to Figure 3 - symmetry in the displacement and polarization responses due to mechanical and electrical point forces can be observed. From Eq. (51), it is observed that $c_2^p = \tilde{c}^p$ which implies that $\mathbf{M}_2(\mathbf{k}) = \tilde{\mathbf{M}}(\mathbf{k})$ from Eq. (55), and this in turn implies that $\mathbf{M}_{\mathbf{uE^0}}(\mathbf{k}) = \mathbf{M}_{\mathbf{Pf}}(\mathbf{k})$ from Eq. (71) and (72). In Figure 5, there is no effect of $l$ on the polarization response as observed from Figures 5(a) and (b); (c) and (d) respectively. As $\delta$ increases, the curves get flatter.

## VI. CONCLUSIONS

A flexoelectric PD theory is laid out. The PD formulation potentially opens up possibilities for a more efficacious modeling of fracture propagation in nanostructures wherein flexoelectric effects could be predominant. In addition to the length scale parameter pertaining to the PD setup, the theory accounts for the higher order gradient length scale parameter and their combined effect on the response. This may perhaps be exploited in nano-transducers for enhanced energy harvesting by an optimal tuning of both the length scale parameters, thereby leading to a better design of such engineering nanostructures. Under static deformation, the proposed model is effectuated in the specific context of an infinite three dimensional body with a point mechanical force and a point electric force. The effect of different length scales on the response is carefully studied and a few observations of interest are drawn thereupon. For instance, even as increasing the PD length scale results in initial softening followed by stiffening, increase in the model length scale associated with strain gradient always results in stiffening. As one moves away from the point of application of the force, displacement under electrical force or polarization under mechanical force are observed to change sign. Also the ratios of displacement to electrical force and polarization to mechanical force are precisely the same.